\documentclass[]{article}
\usepackage[utf8]{inputenc}
\usepackage{xcolor}
\usepackage{tikz}
\usepackage{adjustbox} 
\usepackage{tabularray}

\usepackage{supertabular}
\usepackage{caption}

\usepackage{url}
\usepackage{authblk}

\graphicspath{{Figures/}} 

\usepackage{subcaption}

\usepackage{pifont} 
\newcommand{\checkmark}{\ding{51}}%
\newcommand{\xmark}{\ding{55}}%

\usepackage{titlesec}

\usepackage[resetlabels,labeled]{multibib}
\newcites{Seminar}{Seminar presentation}
\title{Protecting GNSS Against Intentional Interference}
\author[1]{Arul Elango}
\author[1]{Ahmed Al-Tahmeesschi}
\author[1]{Mikko Saukkoriipi}
\author[1]{Titti Malmivirta}
\author[1]{Laura Ruotsalainen}
\affil[1]{Department of Computer Science, University of Helsinki, Finland.}
\date{March 2022, Helsinki, Finland}

\begin{document}


\begin{titlepage}
    \begin{center}
        \vspace*{1cm}
            
        \Huge
        \textbf{WHITE PAPER: \\  Protecting GNSS Against Intentional Interference}
            
        \vspace{1.5cm}
        \Large  
        \textbf{Arul Elango\textsuperscript{1}, Ahmed Al-Tahmeesschi\textsuperscript{1},  Mikko Saukkoriipi\textsuperscript{1},  Titti Malmivirta\textsuperscript{1}, Laura Ruotsalainen\textsuperscript{1}}

        \vspace*{0.25cm}
        \normalsize 
        \textbf{\textsuperscript{1}Department of Computer Science, University of Helsinki, Finland.}
       \vspace*{1cm}
       
       \textbf{March 2022, Helsinki, Finland}

\begin{figure}[!b]
	\centering
	\includegraphics[clip, trim=0.0cm 0.0cm 0.0cm 0.0cm,width=1.\columnwidth]{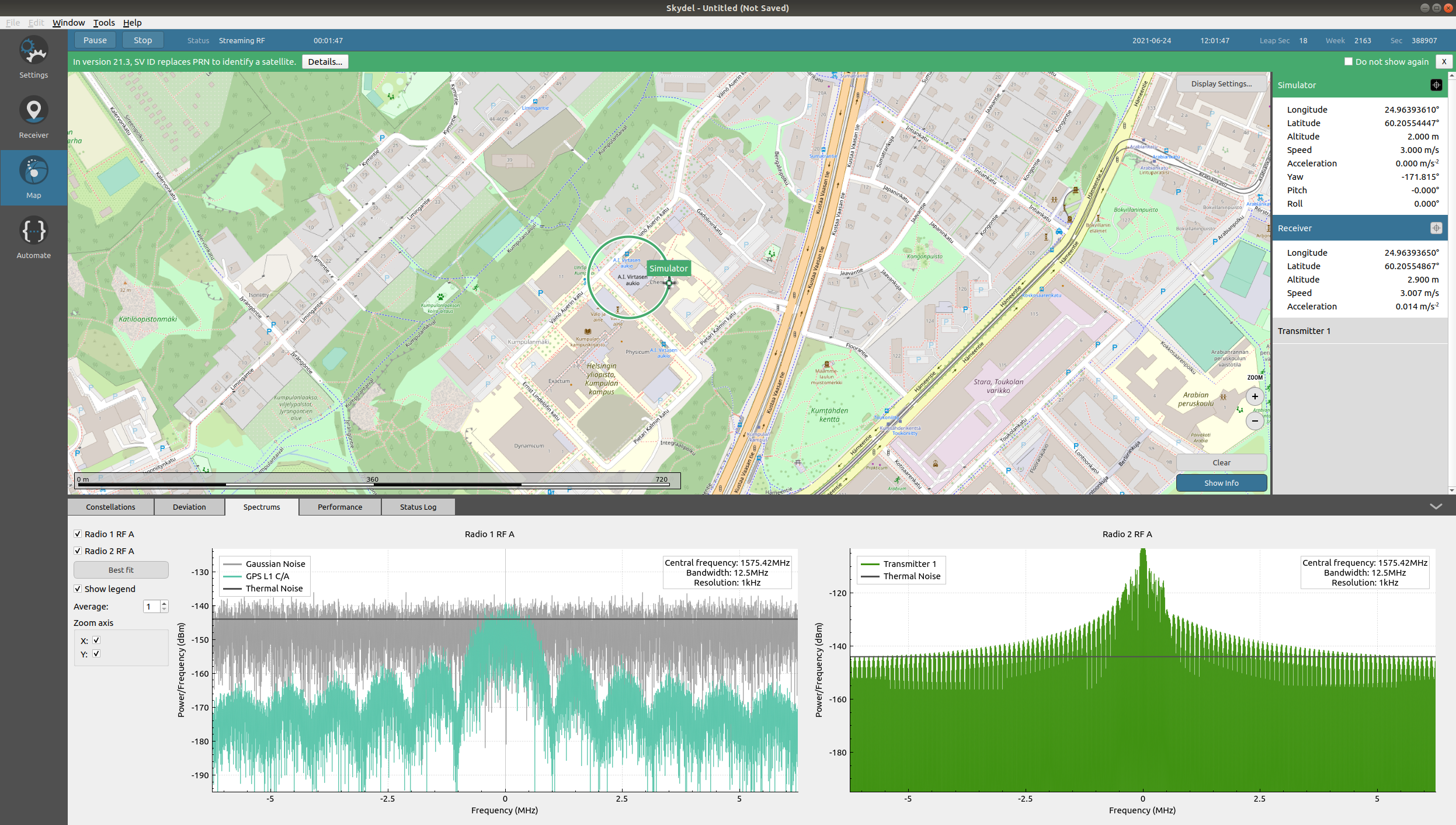}
	\caption*{Photo @Skydel, Orolia}
	\label{fig:cover-photo}
\end{figure}
       
    \end{center}
\end{titlepage}


\maketitle

\begin{abstract}
The vulnerabilities associated with modern systems relying on Global Navigation Satellite Systems (GNSS) due to intentional and unintentional interference is an increasing threat. Since radio frequency interference (RFI) significantly degrades the performance of a GNSS receiver. Several traditional critical applications such as aviation, maritime and rail transport systems to more recent applications such as autonomous vehicles, can be severely affected by such undetected nor mitigated RFIs. Moreover, critical infrastructures such as power supply and money transfer, are becoming more and more dependent on the accurate timing information provided by GNSS. Thus, interference detection and management techniques are crucial to be utilised in order to reduce interference effects. This paper offers a state-of-the-art review of several proposed methods for interference detection and mitigation with solutions ranging from traditional to machine learning-based approaches. In addition, to be able to characterise the RFI threats and develop mitigation techniques, it is essential to monitor RFI systematically and share recorded data with interested entities. Therefore, three GNSS threat monitoring systems are briefly described. This White paper is a compilation of the seminar presentations given at a seminar on "Protecting GNSS against intentional interference" in March 2022 in Finland.

\end{abstract}

\pagebreak

\section{Introduction}

Modern society, especially its critical infrastructure, is becoming more and more reliant on accurate and reliable Position, Velocity and Timing (PVT) information obtained using Global Navigation Satellite Systems (GNSS). GNSS signals travel more than 20000 km from the satellite signal transmitter in space to the user receiver on Earth. Radio waves disperse energy as they propagate; therefore the signals are very weak and vulnerable to interference when they reach the receiver. Interference caused by the atmosphere and signal reflections on the ground have been active research areas already for decades. Intentional interference, namely operations deliberately performed by humans for interfering with GNSS signals, is a continuously increasing problem. Its incidence is anticipated to increase with the increase of number and relevance of the applications relying on GNSS. 
 
 \subsection{Basics of GNSS positioning}
 
GNSS encompass the United States Global Positioning System (GPS), the Russian GLONASS, Chinese COMPASS/BeiDou and the European Galileo systems. In GNSS based positioning the traverse time of a signal from the satellite to the user receiver antenna is estimated. When this time is multiplied by the speed of light a geometric range between the satellite and the user is obtained. In an ideal case, measurements from three satellites would provide an accurate three dimensional position of the user. In reality, the measurements are erroneous, the main error source being the timing errors between the receiver clock and the satellite clock from the system time. Therefore the measured range is called the pseudorange. The satellite clocks are precise and synchronized by the ground control segment of the system. However, the clocks in the user receivers are low-cost with typically a large timing error which has to be estimated as a parameter in the navigation solution. Observations from at least four satellites are needed for three dimensional positioning, namely the fourth observation is used for resolving the receiver clock error. Any electromagnetic source interacting with the signals is interfering with the previously mentioned process of estimating the traverse time. In the atmosphere, mainly in the ionospheric layer, electron concentration causes delays in signal propagation. Signal reflections on the ground, when added to the line-of-sight signals from the satellites, cause self-interference to the signal called multipath. Intentional interference is the most harmful type of interference because it is specifically designed to interfere with GNSS operations and will be discussed a bit in more detail in the following section and throughout this document.  

\begin{table}[]
\caption{List of presentations}
\begin{adjustbox}{width=1\textwidth}
\label{tab:presntations}
\begin{tabular}{|l|l|l|}
\hline
\multicolumn{1}{|c|}{\textbf{Title}}                                                                                                       & \multicolumn{1}{c|}{\textbf{Presenter}} & \multicolumn{1}{c|}{\textbf{Affiliation}} \\ \hline
\begin{tabular}[c]{@{}l@{}}Using the Swedish CORS Network to detect GNSS interference, \\ results from tests and measurements\end{tabular} & Mikael Alexandersson                    & FOI                                       \\ \hline
\begin{tabular}[c]{@{}l@{}}R\&D Activities on Resilient PNT at Finnish Geospatial Research \\ Institute\end{tabular}                       & Zahidul Bhuiyan                         & FGI                                       \\ \hline
PNT\& GNSS Testing Techniques to Mitigate Risk                                                                                             & Robert Burke                            & Orolia                                    \\ \hline
GNSS anomaly monitoring                                                                                                                    & Arul Elango                             & University of Helsinki                    \\ \hline
European Union’s Galileo and its PRS                                                                                                       & Jari Hänninen                           & Traficom                                  \\ \hline
GNSS interference by space weather storms                                                                                                  & Kirsti Kauristie                        & FMI                                       \\ \hline
\begin{tabular}[c]{@{}l@{}}SWEPOS data quality monitoring – GNSS signal interference and \\ disturbance monitoring system\end{tabular}     & Kibrom Ebuy Abraha                      & Lantmäteriet                              \\ \hline
Interference detection, localization, and mitigation in GNSS                                                                               & Elena Simona Lohan                            & Tampere University                        \\ \hline
Jammer fingerprinting                                                                                                                      & Titti Malmivirta                        & Univeristy of Helsinki                    \\ \hline
The ARFIDAAS system design and operation                                                                                                   & Aiden Morrison                          & SINTEF                                    \\ \hline
GNSS – Threats and Countermeasures                                                                                                         & Philipp Richter                         & u-blox                                     \\ \hline
\begin{tabular}[c]{@{}l@{}}Advanced RFI Detection, Analysis and Alerting System II \\ (ARFIDAAS II)\end{tabular}                           & Laura Ruotsalainen                      & University of Helsinki                    \\ \hline
GNSS jamming \& spoofing attacks, techniques, countermeasures                                                                              & Stefan Söderholm                        & Septentrio                                \\ \hline
\begin{tabular}[c]{@{}l@{}}Analysis of a full year of multi-band RFI data from multiple sites \\ in  Europe\end{tabular}                   & Nadia Sokolova                          & SINTEF                                    \\ \hline
\end{tabular}
\end{adjustbox}
\end{table}


\subsection{Intentional Interference}
 
Intentional interference, called either jamming or spoofing depending on the method of implementation and effect on the GNSS signal, deteriorates the obtained PVT solution significantly or completely denies its computation. Jamming means transmission of radio frequency energy at some of the GNSS frequency bands which masks GNSS signals at the related band with noise and thereby prevent the PVT computation \cite{dovis2015gnss}. Although the use of jammers, i.e., equipment used for transmitting a jamming signal, are illegal in most of the countries, many people still use them for the protection of personal privacy. Personal privacy protection in this context means preventing the operation of GNSS receiver that could track its own location and transmit the information to for example the employer monitoring the working time or place \cite{pullen2012gnss}. Jammers transmit signals with power and characteristics disturbing acquisition and tracking of GNSS signals \cite{mitch2011signal}. The use of easy-access low-cost jammers results in degraded positioning accuracy or total loss of GNSS signals and, therefore, may cause serious damage if the jamming signals are not properly detected and their effects mitigated. In addition to privacy protection purposes, intentional interference is caused by political activists with political agenda, cybercriminals for financial purposes and foreign states to damage others' systems (\citeSeminar{ublox_Richter}, presentation). These latter groups doing for criminal purposes have usually higher capabilities and therefore create more significant threat for the GNSS dependent systems. Spoofing, in turn, means the malicious transmission of counterfeit GNSS-like signals and fools the receiver to output an erroneous PVT solution. The greatest danger of spoofing is that its occurrence can be difficult to detect. Fortunately, it is so difficult to be implemented unobtrusively that it cannot be done with off-the-shelf devices without special expertise. In order to be able to create a credible spoofing attack one must be able to generate true GNSS signals including data, modulation and timing, maintain time synchronization close to true GNSS time and adapt the signal power levels to match those of the true signals (\citeSeminar{Soderholm_Septentrio}, presentation).    

It should be noted, that some radio devices, such as amateur radio and military systems, cause natural "jamming" for the GNSS signals (\citeSeminar{Soderholm_Septentrio}, presentation). In addition to interference, GNSS systems can fail such as Galileo ground system atomic clock failure in December 2020, have created interference to GNSS (\citeSeminar{ublox_Richter}, presentation). 
 
 \subsection{Contents and Organization of this White Paper}
In order to secure the critical infrastructure and assure continuous availability of GNSS, all forms of  interference have to be detected and their impact on GNSS-based systems mitigated. This White Paper has been compiled based on presentations given at the "Protecting GNSS against intentional interference" seminar, held on 24.3.2022 at the University of Helsinki. The presentations discussed the state-of-the-art activities related to GNSS vulnerabilities in Nordic countries.
 
The impact and detection of interference has been discussed widely in the literature, e.g. \cite{Guo_2021} for ionospheric effects, \cite{Borio_2013} for jamming and \cite{psiaki2014gnss} for spoofing. However, current detection methods are not sufficient for the ever-evolving interference attacks and society's growing dependence on GNSS. Therefore, Section 2 of this document will look into most recent methods for detecting the interference, unintentional (Section 2.1) and intentional (Section 2.2). Section 2.2 will also go through methods for identifying the specific jamming device in order to encounter its use. Countermeasures to protect the critical infrastructure will be discussed in Section 3 and the deployment of large scale interference monitoring infrastructure in Section 4. The document will summarize the main findings in Section 5. 

\vspace{\baselineskip} 




\section{Interference detection}

A GNSS radio receiver usually consists of two main parts: the analog and the digital part. The analog part includes the antenna and the radio front-end (RFE). The RFE is responsible for amplifying the weak signal received by the antenna, filtering it by a band-pass filter to minimize out-of-band contributions and down-converting it to Intermediate Frequency (IF). Finally, the Analog-to-Digital Converter (ADC) converts the analog signal to a digital version, with an Automatic Gain Control (AGC) lower quantization losses apparent in the digitization process by automatically adjusting the signal dynamics.

\begin{figure}[htp]
    \centering
    \includegraphics[width=12cm]{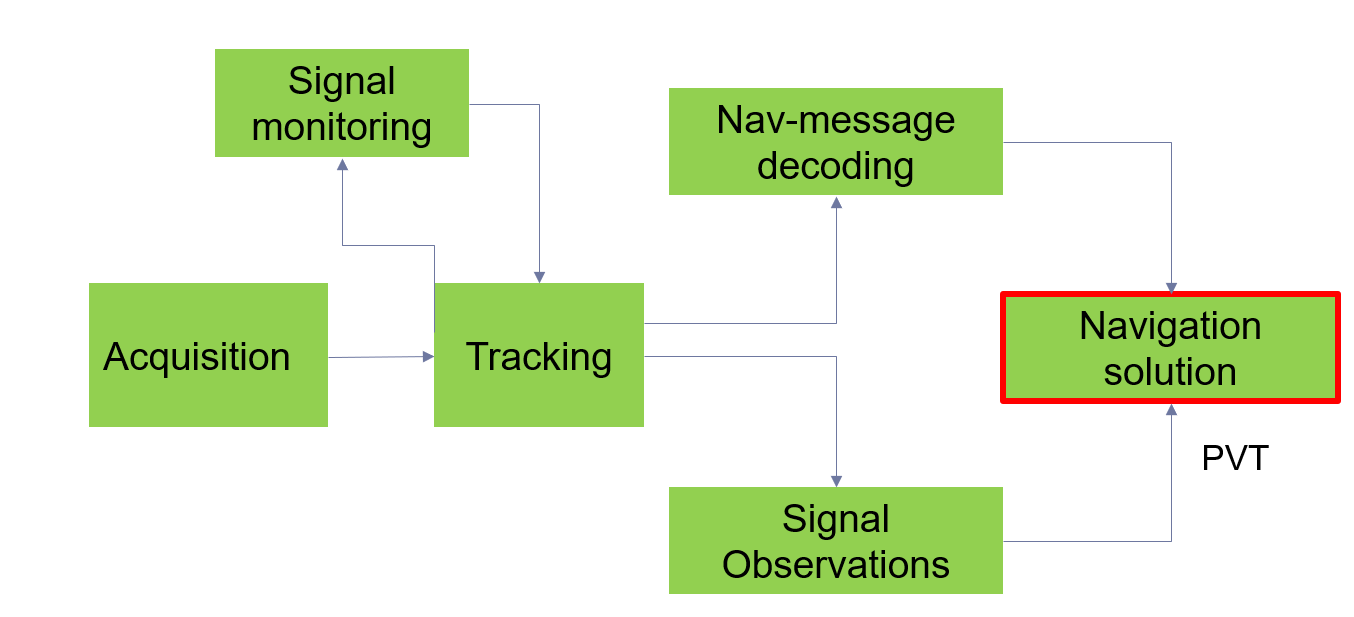}
    \caption{GNSS receiver's processing blocks}
    \label{Rx}
\end{figure}

The digital part of the GNSS receiver is processing the digitized signal to obtain a PVT solution. Figure \ref{Rx} shows the GNSS receiver's digital processing blocks. Acquisition block detects the signal presence and gives a rough estimate of the two parameters required in the process, code delay and Doppler shift. When these are obtained, processing transfers to the tracking stage, where the code delay and Doppler shift estimates are fined down. Tracking outputs are monitored by the signal monitoring block, which also computes the signal quality measures, Carrier-to-Noise ($C/N_0$) values. Tracking provides signal observations called carrier phases and pseudoranges, which together with the navigation data information (satellite parameters and time of the signal transmission) packed into the signal, are used for computing the PVT solution at the navigation block.

\subsection{Detecting unintentional interference}

As discussed before, concentration of free electrons in the atmosphere, specifically ionosphere, degrade the GNSS signal and therefore their effect has to be compensated for. When the positioning solution may be computed using signals at more than one GNSS frequency band, the ionospheric effect can be cancelled out, and in the case of only single frequency it can be modelled and its impact can be reduced. However, sometimes existing electron density irregularities disrupt the signal propagation and introduce scintillation, which means fluctuating signal's amplitude and phase \cite{dovis2015gnss}. The extent and effect of scintillation is dependent on solar and geomagnetic activity, geographical location, time and signal frequency. For example day and night variation on the electron density may be 100 fold, and 10 fold according to solar cycle (\citeSeminar{FMI_Kauristie}, presentation). The most feasible method for detecting GNSS scintillation is the use of a dense GNSS receiver network, where the network GNSS stations are static and provide scintillation indices and other observables indicating the signal quality. Some networks provide also near-real-time data, for both research and operational use.  

\subsection{Detecting intentional interference}

\begin{figure}[htp]
    \centering
    \includegraphics[width=12cm]{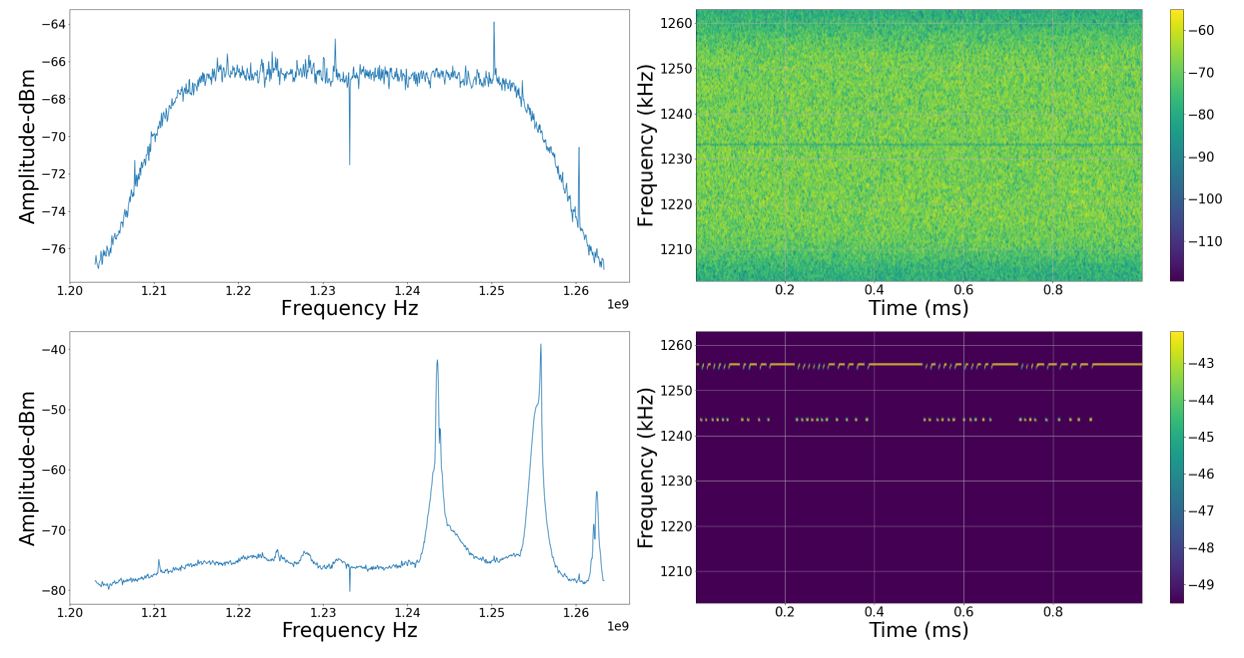}
    \caption{Original GNSS signal (top) and jammed signal (bottom) spectrum (left) and time-frequency (right) plots (\protect\citeSeminar{Arul_Elango_UoH}, presentation).}
    \label{fig:Methodological approaches in dealing with interference}
\end{figure}

Jamming deteriorates the GNSS signal. Figure \ref{fig:Methodological approaches in dealing with interference} shows the spectrum (left) of the original GNSS signal (top) and jammed signal (bottom) and time-frequency plots for the signals (right), respectively. Detection of the presence of a jamming signal is crucial for being able to protect the GNSS signal reception and provision of the PVT solution. Interference detection methods can be classified into domain-specific techniques and generic techniques. Domain-specific refers to implementation of methods in specific receiver stage (front-end, acquisition, tracking or navigation techniques) (\citeSeminar{Lohan_Tre_Univ}, presentation). The front-end specific methods include for example monitoring the AGC level changes, acquisition looking at time-based and frequency-based transforms, tracking monitoring the $C/N_0$ levels and navigation correlation of propagation-dependent observables and consistency checks With additional sensors. Generic techniques include techniques like using machine learning for anomaly detection or radio frequency fingerprinting. 

\vspace{\baselineskip} 

\subsubsection{Jammer fingerprinting}
Radio frequency fingerprinting refers to identifying radio transmitters based on unique signal distortions. The distortions are caused by irregularities in hardware components in signal processing chains. The distortions are unique to each device, which makes it possible to use radio frequency fingerprinting to detect spoofing (\citeSeminar{Lohan_Tre_Univ}, presentation) as well as identify individual jamming devices. While commonly used cheap jammers tend to not have very complex signal processing chains, the signals from similar jammer devices can and do still vary. Machine learning has provided good performance in classifying individual jammers (\citeSeminar{Malmivirta_HY}. 

\begin{figure}[htp]
    \centering
    \includegraphics[width=12cm]{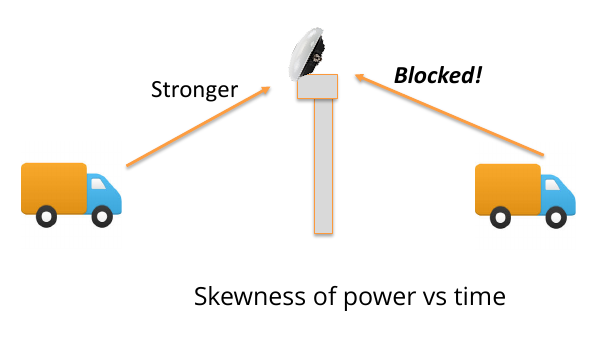}
    \caption{Interference source inside truck approaching antenna from left (\protect\citeSeminar{Soderholm_Septentrio}, presentation). The jamming signal gets stronger when the source approaches the antenna and decreases fast when a tilted choke-ring antenna blocking the signals from below is used.}
    \label{fig:soderholm-jammer-direction}
\end{figure}

In addition to identifying the sources of interference, it is also important to be able to localize them or find the signal direction in order to stop such harmful activities. Multi-antenna arrays can be used for localization or direction finding (\citeSeminar{Lohan_Tre_Univ}, presentation). An example of jammer direction sensing was presented by Stefan Söderholm (\citeSeminar{Soderholm_Septentrio}, presentation). Dual-antenna receiver was placed on a highway with several lanes. Direction detection method used was based on power changes in signals as a function of time. This was done by directing one antenna to face left figure \ref{fig:soderholm-jammer-direction} so that it received stronger signals when interference source was approaching from left. After the signal source has passed the antenna the signal power level decreases rapidly. This is due to the use of a tilted professional grade choke-ring antenna, which is implemented in a way that it blocks the signals arriving from the below the antenna. However, as the antenna is tilted, the "below" is actually pointing to right of the figure.

\begin{figure}
    \centering
    \includegraphics[width=10cm]{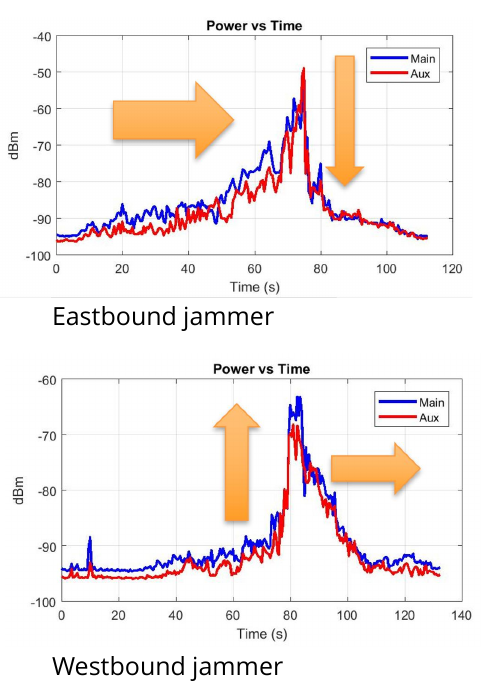}
    \caption{Jammer direction detection: signal power level changes as function of time as interference source passes antennas (\protect\citeSeminar{Soderholm_Septentrio}, presentation). Up: jamming source is arriving from left, assigned here as the east direction and the power level increasing while approaching the antenna at the time 75 s and decreasing rapidly. Down: similar phenomena may be observed while the jamming source is approaching the same antenna setup from west. }
    \label{fig:soderholm-power-levels}
\end{figure}

Direction of the source can be seen from power changes in signals as seen in figure \ref{fig:soderholm-power-levels}. Upper graph in figure \ref{fig:soderholm-power-levels} shows the signal power as a function of time when interference source is carried by a truck arriving from the left, where we consider the left side of the figure being east direction. The figure shows that when the interference source approaches the antenna, the signal power level rises slowly.  Same can be seen in lower graph of figure \ref{fig:soderholm-power-levels} plotting the change of the jamming signal power level when the source is westbound.



\vspace{\baselineskip} 

\vspace{\baselineskip} 



\subsection{GNSS anomaly detection}
Anomaly means different or abnormal, in GNSS domain we can consider anomalies all features in the signal that are caused by a signal degradation source. Thereby, GNSS anomaly detection means the detection of both unintentional and intentional interference. The goal is to both observe that something is wrong with the signal and in the best case to classify its source. At present, anomaly detection provides promising results when machine learning and especially deep learning based methods are used. 

The conventional deep learning methods, usually based on neural networks, ignore the temporal domain, which is not very feasible in the case of navigation data. Long Short Term Memory (LSTM) methods are so called recurrent models, which process the data as time sequences. An autoencoder is an unsupervised neural network able to learn efficient codings of unlabelled data. It learns a representation for the input data by training the neural network to ignore noise. Autoencoder is comprised of an encoder function, which converts the input data into a diﬀerent representation, and a decoder function, which converts the new representation back into the original format. Autoencoders may be realized with various neural network architectures. To accommodate for temporal correlations in the data, the encoder and decoder are implemented using the LSTM architecture. In anomaly detection, encoder section first modifies the data into a time-dependent sequence, preserving only the most important features of the data. The decoder then modifies the data back to the original format. This back-and-forth editing will result in an error that will be minimized. The training data is clean, meaning it has no anomalies. If a beforehand set threshold is exceeded when the algorithm uses real data, anomalies are declared.

In general, part of the GNSS signal is considered to be anomalous in comparison to the whole GNSS signal \cite{Wang_2018_Machine_Learning}.
In addition, for the time domain representation of the GNSS signal, signal behaviour changes and cannot be easily visualized. Therefore, frequency domain representation is utilized for anomaly detection (\citeSeminar{Arul_Elango_UoH}, presentation). The method which is specific to deal with time series is the LSTM which falls in the category of supervised deep learning method \cite{Farshchi_2018}.

\begin{figure}[htp]
    \centering
    \includegraphics[width=12cm]{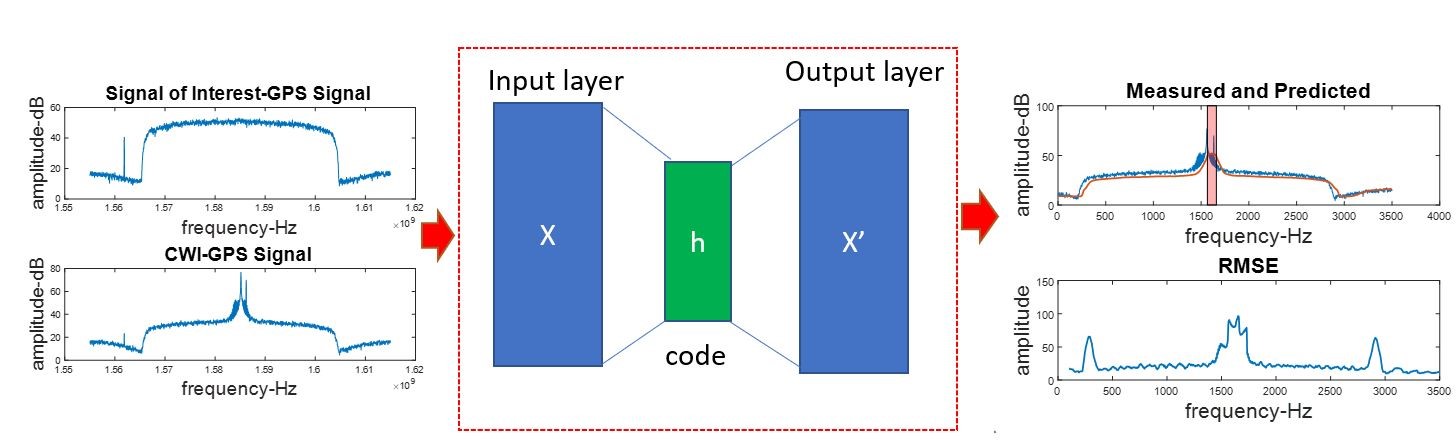}
    \caption{LSTM Autoencoder based GNSS Anomaly detection (\protect\citeSeminar{Arul_Elango_UoH}, presentation).}
    \label{fig:LSTM Autoencoder based GNSS Anomaly detection}
\end{figure}

An autoencoder network comprises of two parts: an input layer (endcoder) and an output layer (decoder). The encoder block compresses the N-dimensional GNSS data into an X-dimensional code (where X $<$ N), which contains most of the features carried in the input. The decoder,  reconstructs the input from the lower-dimensional code (also refereed to in the literature as latent space representation). The way one can use trained autoencoders for anomaly detection is that in normal conditions, when normal signal is fed into the network, it can only reconstruct the signal that does not include abnormalities. If the error between the original signal and the reconstructed signal is small then the data containing anomalies can be used to recognize the changes in the original signal. 
In figure \ref{fig:LSTM Autoencoder based GNSS Anomaly detection} initially the signal of interest (GPS signal) is fed to the network so that the output layer could be able to reproduce the same GPS signal but when jamming signal is included  (continuous wave interference (CWI) in our case) the root mean square error (RMSE) between the input and the output is calculated if that error exceeds the predefined threshold then it is spotted as anomaly region as indicated in red coloured region at the top right hand side of figure \ref{fig:LSTM Autoencoder based GNSS Anomaly detection}.  

GNSS interference research and development requires the use of simulated data. Many of the interference events are rare and difficult to be captured from the real signal data, and as the research is still actively ongoing their identification and correct classification might not be doable. Orolia Skydel GNSS signal simulator (\citeSeminar{OROLIA_Robert_Burke}, presentation) provides not only simulating the different kinds of jamming, spoofing and multipath signals but also support for multiple technologies such as GNSS Augmentations, RTK, Low Earth Orbiting (LEO) Satellite simulations. Orolia Skydel uses Graphical Processing Unit (GPU) accelerated computing to create GNSS/RF signals digitally and software defined radios (SDR) to output the RF. Another salient feature of this simulator is that it provides an intuitive application program interface (API) to configure the simulator. 

\section{Counter measures for interference}

GNSS interference mitigation techniques can be roughly divided into five categories; signal processing means \cite{Ferrara_2018_narrowband_interference} antenna configuration-based methods, encryption of signals, sensor integration, and system deployment related approaches. This section discusses the four first categories, system deployment related approaches, namely interference monitoring infrastructures, are discussed in the following section.

\subsection{Signal processing based interference mitigation}
Signal processing based interference mitigation may be grouped into frequency-domain and time-domain techniques \cite{dovis2015gnss}. Interference mitigation using signal processing means, namely filtering, is best to be performed in the frequency-domain. Two of the most feasible filtering methods are called Notch and Frequency Domain Adapt Filtering (FDAF) \cite{Falletti}. They are both able to adjust at some extent to the situation, where the interfering signal changes its spectral characteristics in time. Although the processing of the signal in the time-domain is feasible for the interference detection purposes, it often is not for mitigation as the interfering signal is usually mixed with the original signal.  

\subsection{Antenna configuration-based mitigation}
Antenna polarization defines the geometrical orientation of the oscillations of the wave. To maximize the received signal strength, the polarization of the receiving GNSS antenna must match the transmitted signal. GPS satellites transmit signals in the form of Right-Hand Circularly Polarized (RHCP) wave and the polarization may change from right to left when the signal is reflected from obstacles. As a result, multipath effects will be diminished in the pre-processing stage. The observation of GPS signal using right hand and left-hand circular polarization antennas will help us better understand the characteristics of multipath GNSS signals and thus mitigate their effects \cite{Dinesh_2004}. This is achieved by introducing left hand polarization to handle NLOS signals to nullify the effects of multipath. Nevertheless, the mitigation effectiveness on satellite visibility and may not be enough for treating all the kind of threats. In the case of jamming, adaptive antenna systems and null steering (the use of spatial signal processing for nulling the interference) antennas provide mitigation (\citeSeminar{ublox_Richter}, presentation). Spoofing mitigation may be done by using antenna arrays for angle of arrival (AoA) detection and thereby evaluating if the signals are really coming from the space or falsely from the ground.  

\subsection{European Union's Galileo}

New GNSS systems provide system based means for interference mitigation. The best example of such is Europe's own Galileo system's Public Regulated Service (PRS) which will provide the European authorities protection against GNSS spoofing.

Galileo is an entirely civilian global navigation system managed by the European Union (EU) and the European Space Agency (ESA). In the management of Galileo, the EU is responsible for the legal and political issues, and ESA manages the technological development. Galileo has been designed to be open, global, and highly reliable while being completely independent of the other satellite systems. The Galileo project contracts were signed in 1999, and the declaration of Galileo's initial service happened in December 2016. While Galileo is an independent satellite navigation system, it was not designed to replace or compete with other global navigation systems. Instead, it is designed to be interoperable with them while offering complete sovereignty.

Once Galileo is fully operational, it will offer four services \cite{Galileo-services}. These are Open Service (OS), High Accuracy Service (HAS), Public Regulated Service (PRS), and the Search and Rescue Service (SAR). OS is the conventional navigation service provided by Galileo globally and free of charge and directed at mass markets requiring meter level accuracy. HAS, on the other hand, is aimed at professional or commercial applications requiring centimeter level performance, which is higher than offered by the OS. It will be offered globally and free of charge. 

SAR/Galileo is EU's contribution to Cospas-Sarsat, an international satellite-based search and rescue distress alert detection system. SAR/Galileo is two-way, meaning it forwards the distress signals to the relevant rescue authorities and returns an automated acknowledgement message back to the user in distress \cite{Sar}. The carrier frequencies of OS, HAS and PRS are introduced in table \ref{tab:galileo_frequency}. High precision and reliable timing are critical in satellite navigation. To meet these requirement, each Galileo satellite has four high-precision atomic clocks. Two of these are Rubidium Atomic Frequency Standard (RAFS) clocks with a precision of 10 ns per day. The other two are Passive Hydrogen Maser (PHM) clocks with a precision of 1 ns per day. The RAFS clocks are used to secure short-term stability, while the PHM clocks guarantee short and long-term stability \cite{hofmann_2007_gnss}.

Even though Galileo is relatively new, and the first Galileo-enabled mobile phone came to market in 2016, it is already widely used in civilian applications. Currently there are over 3 billion Galileo-enabled mobile phones and over 15 million Galileo-equipped cars (\citeSeminar{hanninen_traficom}, updated post presentation).

\begin{table}[h]
\caption{Galileo signal lexicon and signals used by different Galileo services.}
\label{tab:galileo_frequency}
\begin{center}
\begin{tabular}{|c|c|c|c|c|}
\hline
\begin{tabular}{@{}c@{}}Frequency \\ Band \end{tabular} & \begin{tabular}{@{}c@{}}Carrier Frequency \\ (MHz)\end{tabular} & OS & HAS & PRS \\ 
\hline 
E1 & 1575.420 & \checkmark &\xmark  & \checkmark \\ 
\hline
E5 & 1191.795 &  \checkmark & \xmark & \xmark \\ 
\hline
E6 & 1278.750 &  \xmark & \checkmark & \checkmark \\ 
\hline
\end{tabular}
\end{center}
\end{table}

\subsubsection{Galileo Public Regulated Service (PRS)}

PRS is an encrypted navigation service for government-authorised users in the EU and in duly authorised third countries and international organisations. In Finland, the Finnish Transport and Communications Agency (Traficom) is the competent PRS authority (CPA), and preparations for building a national PRS management infrastructure are underway (\citeSeminar{hanninen_traficom}, presentation).

When the PRS reaches full operational capability, it is expected to serve a wide range of authorised users. These users include police, defence forces, border guard, emergency services, and selected security-of-supply companies that are part of the critical national infrastructure and logistics (\citeSeminar{hanninen_traficom}, presentation). When comparing Galileo PRS to open and commercial GNSS services, PRS offers several benefits. Most importantly, it gives protection against interference, jamming, and spoofing, making it significantly more reliable under adversarial conditions. PRS uses two spectrally separated and encrypted carrier frequency bands (E1 \& E6) to maximize the inference resistance and make it more difficult to attack the signal. EU \cite{Law5} also emphasises that PRS must have service continuity even in the most serious crisis situations (\citeSeminar{hanninen_traficom}, presentation).


\subsection{Deeply-coupling Methods} 

Deeply-coupled GNSS receiver and self-contained sensor integration for jamming and spoofing mitigation has not been discussed in the extent it deserves, because it seems to be one of the most promising countermeasures. Sensors are not affected by radio interference and the attitude and acceleration information they provide are good complements to GNSS from other aspects too. An advanced integration method called deep-coupling uses information obtained from self-contained sensors to aid the signal processing algorithms and therefore enhances the robustness of GNSS to interference \cite{Ruotsalainen_2014_Impact}. However, self-contained sensors suffer from biases and drift errors that may decrease the position accuracy substantially. Especially when using consumer grade Micro-Electro-Mechanical System (MEMS) sensors, suited when weight and cost have to be considered.

Typically, deep-coupling with GNSS signals corrects the above-mentioned bias and drift errors in self-contained sensors, but if the interference continues for a longer time forestalling the use of proper GNSS signals, degradation of the combined position solution will incur after some period. Therefore, other sensors or methods have to be used to calibrate the errors in self-contained sensors frequently for a robust result. One emerging technology for error calibration is the use of visual perception. Visual sensors, i.e., cameras, are feasible instruments for constricting the growth of the errors and are also resistant to GNSS interference. Deeply-coupled GNSS, inertial navigation system (INS) and vision fusion has proven to be a feasible method for interference mitigation Integration \cite{Cristodaro_Ultra_Tight_Integration}.

\section{Interference monitoring infrastructure }

The vulnerability in GNSS-based systems to intentional and unintentional interference requires having disparate monitoring systems continuously monitoring the GNSS signals for anomalies. 
In general, monitoring systems can be classified into two main categories; fixed and portable \cite{thombre_2018_gnss}. The fixed monitoring systems are installed at a site to operate over long periods. In addition, the fixed monitoring systems can be expanded to multiple monitoring nodes connected to a central control unit. The raw observed data are sent via the communication channels to the control unit.
The control centre stores the observed GNSS received signals and makes them available for post-processing to the end users. The portable monitoring systems are typically small handheld or mounted on vehicles which are intended to be operated over shorter periods. 



In this section, three fixed European monitoring infrastructures to monitor interference will be explained. Table \ref{tab:monitoring_systmes} shows a comparison summary which is based on the capabilities of the threat monitoring systems; based on their capabilities to classify the interference type, detection of signal spooﬁng, if they are capable of geo-localisation of the interference source, price per station and redeployment capability (i.e., more feasible to change monitoring station position).


\begin{table}[]
\caption{Overview of the considered GNSS threat monitoring systems.}
\label{tab:monitoring_systmes}
\begin{adjustbox}{width=1\textwidth}
\begin{tabular}{|l|l|c|c|c|c|c|c|}
\hline
\begin{tabular}[c]{@{}l@{}}\textbf{Monitoring} \\ \textbf{System}\end{tabular} & \begin{tabular}[c]{@{}l@{}}\textbf{Constellations} \\ \textbf{monitored}\end{tabular}            & \begin{tabular}[c]{@{}c@{}}\textbf{Interference} \\ \textbf{type} \\ \textbf{classification}\end{tabular} & \begin{tabular}[c]{@{}c@{}}\textbf{Spoofing} \\ \textbf{detection}\end{tabular} & \begin{tabular}[c]{@{}c@{}}\textbf{Interference} \\ \textbf{localisation}\end{tabular} & \begin{tabular}[c]{@{}c@{}}\textbf{Interference} \\ \textbf{localisation}\end{tabular} & \begin{tabular}[c]{@{}c@{}}\textbf{Low-price} \end{tabular}& \begin{tabular}[c]{@{}c@{}}\textbf{Redeployment}\end{tabular} \\

\hline
FinnRef                                                      & \begin{tabular}[c]{@{}l@{}}GPS, \\ GLONASS, \\ Galileo, \\ BeiDou\end{tabular} & \xmark                                                                    & \xmark                                         & \xmark                                                & \checkmark      & \xmark  &  \xmark                             \\ \hline
SWEPOS                                                       & \begin{tabular}[c]{@{}l@{}}GPS, \\ GLONASS, \\ Galileo, \\ BeiDou\end{tabular} &  \checkmark                                                      & \checkmark                                     & \checkmark                                                & \checkmark      & \xmark  &  \xmark                             \\ \hline
ARFIDAAS                                                      & \begin{tabular}[c]{@{}l@{}}GPS, \\ GLONASS,\\ Galileo, \\ BeiDou\end{tabular}  & \ \checkmark                                                      & \checkmark                                     & \checkmark                                            & \checkmark            & \checkmark & \checkmark                      \\ \hline
\end{tabular}
\end{adjustbox}
\end{table}

\subsection{Finnish National Reference Network (FinnRef)}
\begin{figure*}[!t]
	\centering
	\includegraphics[clip, trim=0.0cm 0.0cm 0.0cm 0.0cm,width=0.75\columnwidth]{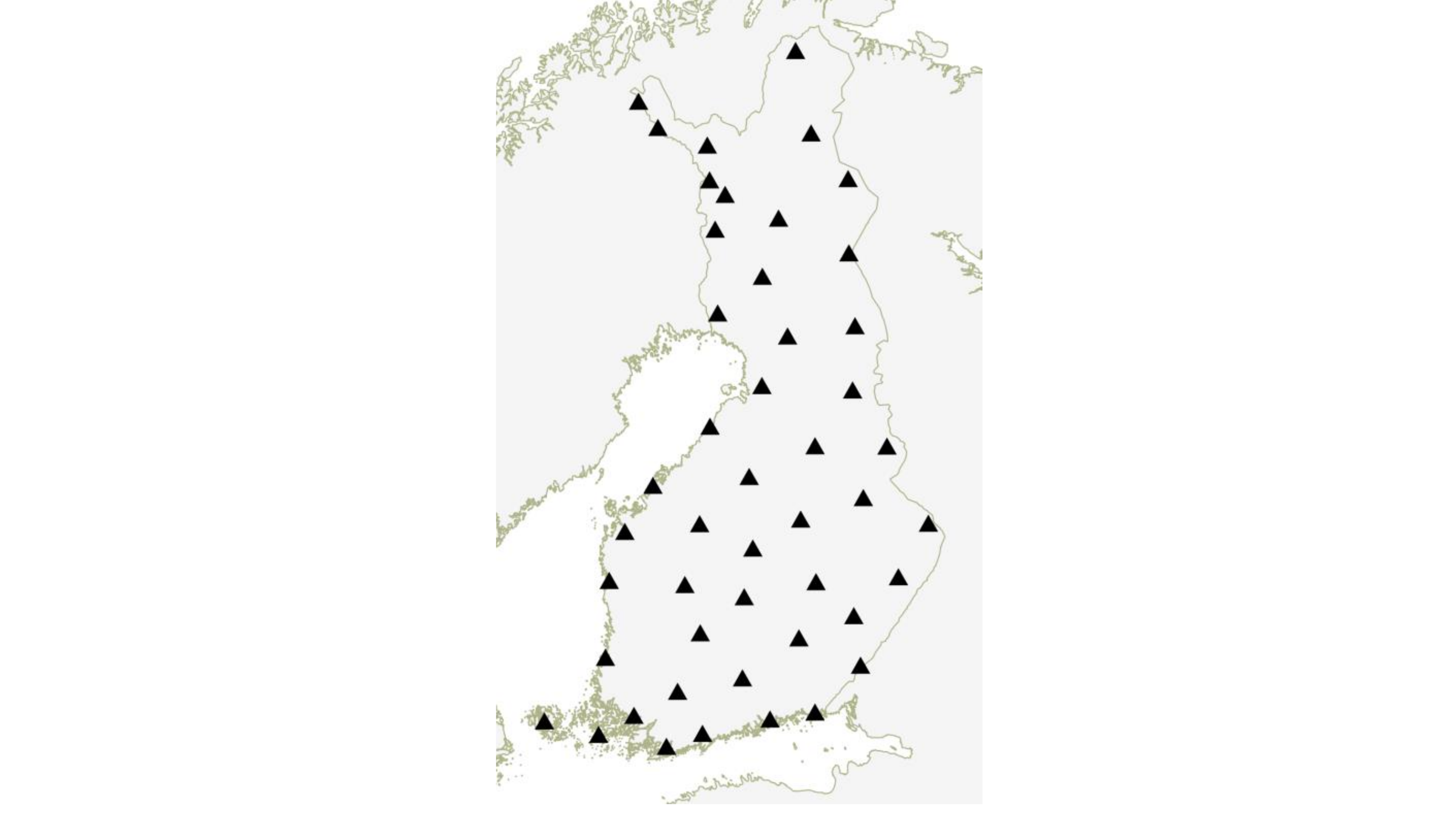}
	\caption{FinnRefs stations' locations (\protect\citeSeminar{Zahidul_FGI}, presentation).}
	\label{fig:finnref2}
\end{figure*}

\begin{figure*}[!t]
	\centering
	\includegraphics[clip, trim=0.0cm 0.0cm 0.0cm 0.0cm,width=0.75\columnwidth]{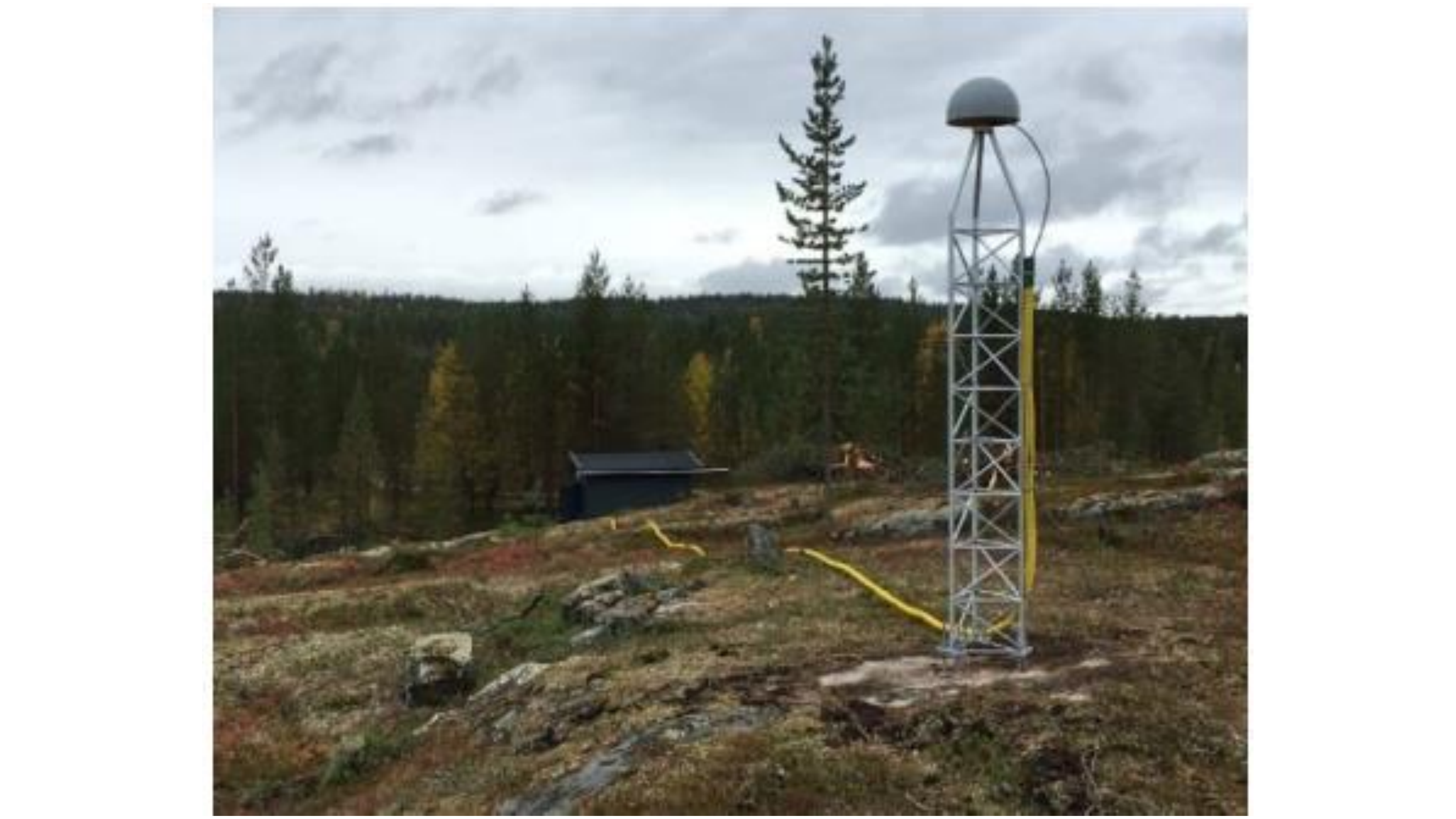}
	\caption{Typical FinnRef station (\protect\citeSeminar{Zahidul_FGI}, presentation).}
	\label{fig:finnref1}
\end{figure*}

Finnish National Reference Network (FinnRef) includes 47 fixed stations to monitor GNSS signal quality with locations as can be seen in figure \ref{fig:finnref2} and a typical station is seen in figure \ref{fig:finnref1}. FinnRef forms the basis for the national reference frame, EUREF-FIN and is maintained by the National Land Survey of Finland. Moreover, some of the stations also serve as International GNSS Service (IGS) stations to provide open access, high-quality GNSS data for scientific, educational, and commercial applications. In addition, one station is co-located with European Geostationary Navigation Overlay Service (EGNOS) Ranging and Integrity Monitoring Station (RIMS) which are used to improve the performance of GNSS. The Real-time positioning service ‘FINPOS’ uses FinnRef data to provide Differential-GNSS (DGNSS) and Network RTK measurement data to improve localisation accuracy for end users. The information from multiple global constellations i.e., GPS, Galileo, GLONASS and BeiDou are monitored. The monitored information is available in RINEX format and in real-time streams for the monitored constellations, where RINEX is a standard format that allows the management and disposal of the measured data (\citeSeminar{Zahidul_FGI}, presentation). In addition, the FinnRef monitors the GNSS signals for interference and jamming. It sends alert messages to Traficom when signal anomalies are detected \cite{GnsS_Finland_Service}. 

\subsubsection{GNSS-Finland Service}
The data obtained from FinnRef forms the backbone for GNSS-Finland Service where the obtained data from the network are continuously analysed in real-time. The service creates a signal quality indicator for the monitored GNSS signal strengths. The indicators are "good", "satisfactory" or "poor". The specific observed GNSS signal at a specific station is estimated as the average $C/N_0$ demonstrating the quality of the signal for all observed satellites excluding satellites with an elevation angle less than 15 degrees. As for the signal quality indicator of the above-mentioned grades based on average $C/N_0$, two thresholds are empirically estimated for all signals. More details on thresholds values can be found at \cite{GnsS_Finland_Service}. The Finnish Meteorological Institute uses the FinnRef receiver data in its space weather services to monitor ionospheric activity.


\subsection{SWEPOS}
The Swedish network of permanent reference stations for GNSS includes 500+ fixed nodes scattered all around Sweden. The receiver nodes are connected to a control centre that provides access to the whole accumulated GNSS data in real-time \cite{Norin2008SWEPOS}. Each node is equipped with antennas and receivers which enable the network to track GPS, GLONASS, Galileo and Beidou signals. In addition to spectrum monitoring, SWEPOS also provides a range of applications such as: network real-time kinematic (NRTK) correction for real-time applications,  Data for geoscientific and meteorological research and serves as the backbone of the Swedish national geodetic reference frame (\citeSeminar{Abraha_Lantmateriet}, presentation). 

\begin{figure*}[!t]
	\centering
	\includegraphics[clip, trim=0.0cm 1.0cm 0.0cm 3.0cm,width=1\columnwidth]{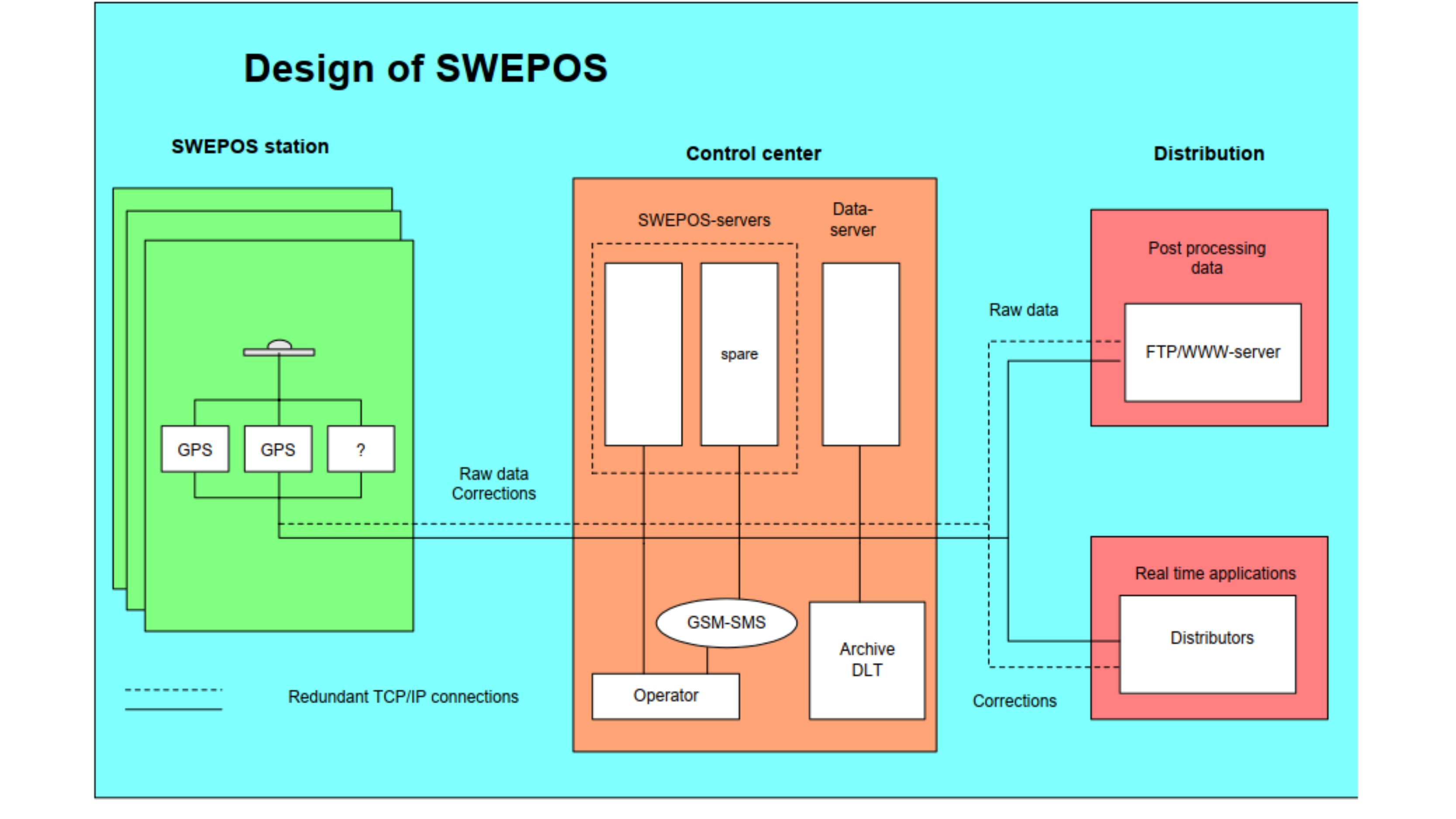}
	\caption{Design of SWEPOS \cite{Hedling2001SWEPOS}.}
	\label{fig:Design_of_SWEPOS}
\end{figure*}

Figure \ref{fig:Design_of_SWEPOS} shows the design of the the SWEPOS-network. The SWEPOS stations are connected to the control centre at the National Land Survey through TCP/IP connection. The quality of raw data and the Differential GPS (DGPS) corrections checked at the control centre, and then, the control centre provides the data in a RINEX format for post-processing through a WWW/FTP-server. In general, most of the SWEPOS stations are built in areas with an undisturbed line of sight to the monitored satellite constellation. 


\begin{figure*}[!t]
	\centering
	\includegraphics[clip, trim=2.0cm 0.0cm 0.0cm 0.0cm,width=1.2\columnwidth]{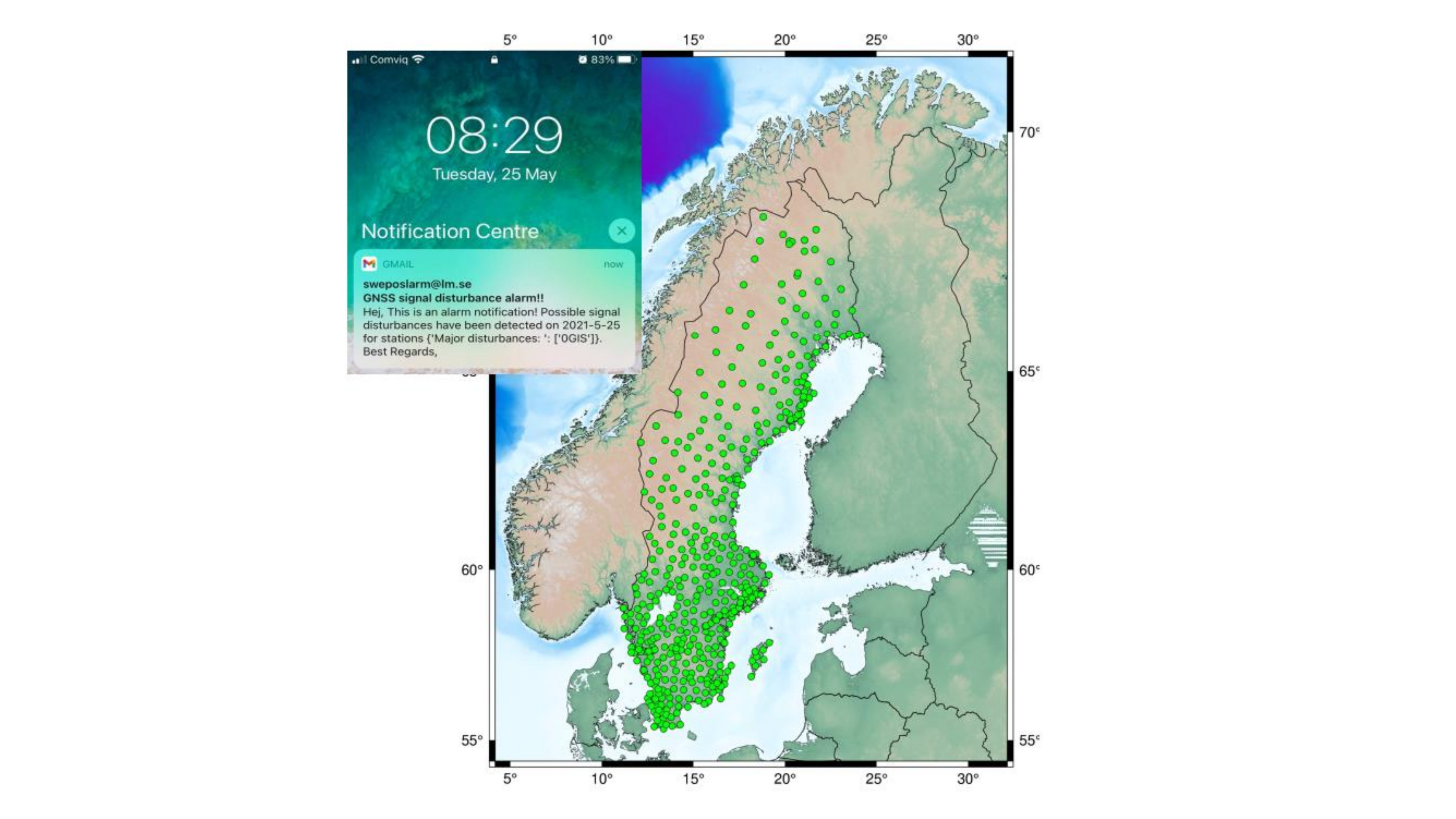}
	\caption{SWEPOS network (green dots) and message alerting system (\protect\citeSeminar{Abraha_Lantmateriet}, presentation).}
	\label{fig:Measurement_system}
\end{figure*}

The SWEPOS network infrastructure is utilised to monitor multi-GNSS constellations
at multi-frequency bands. In addition, testing has been done in order to characterise and detect radio frequency interference in Sweden for both jamming and spoofing events
in near-real-time from $C/N_0$ history and characteristics \citeSeminar{FOI_Alexandersson}. Moreover, it sends alarms when radio frequency interfernece (RFI) is detected (Figure \ref{fig:Measurement_system}. Also, it characterises the GNSS RFI events and geolocates the RFI sources for the use of the Swedish Post and Telecom Authority (PTS) based on the power and duration of the interference (\citeSeminar{Abraha_Lantmateriet}, presentation).



\subsection{Advanced RFI Detection, Analysis and Alerting System  (ARFIDAAS II)}
\begin{figure*}[!t]
	\centering
	\includegraphics[clip, trim=0.0cm 3.5cm 0.0cm 3.5cm,width=1\columnwidth]{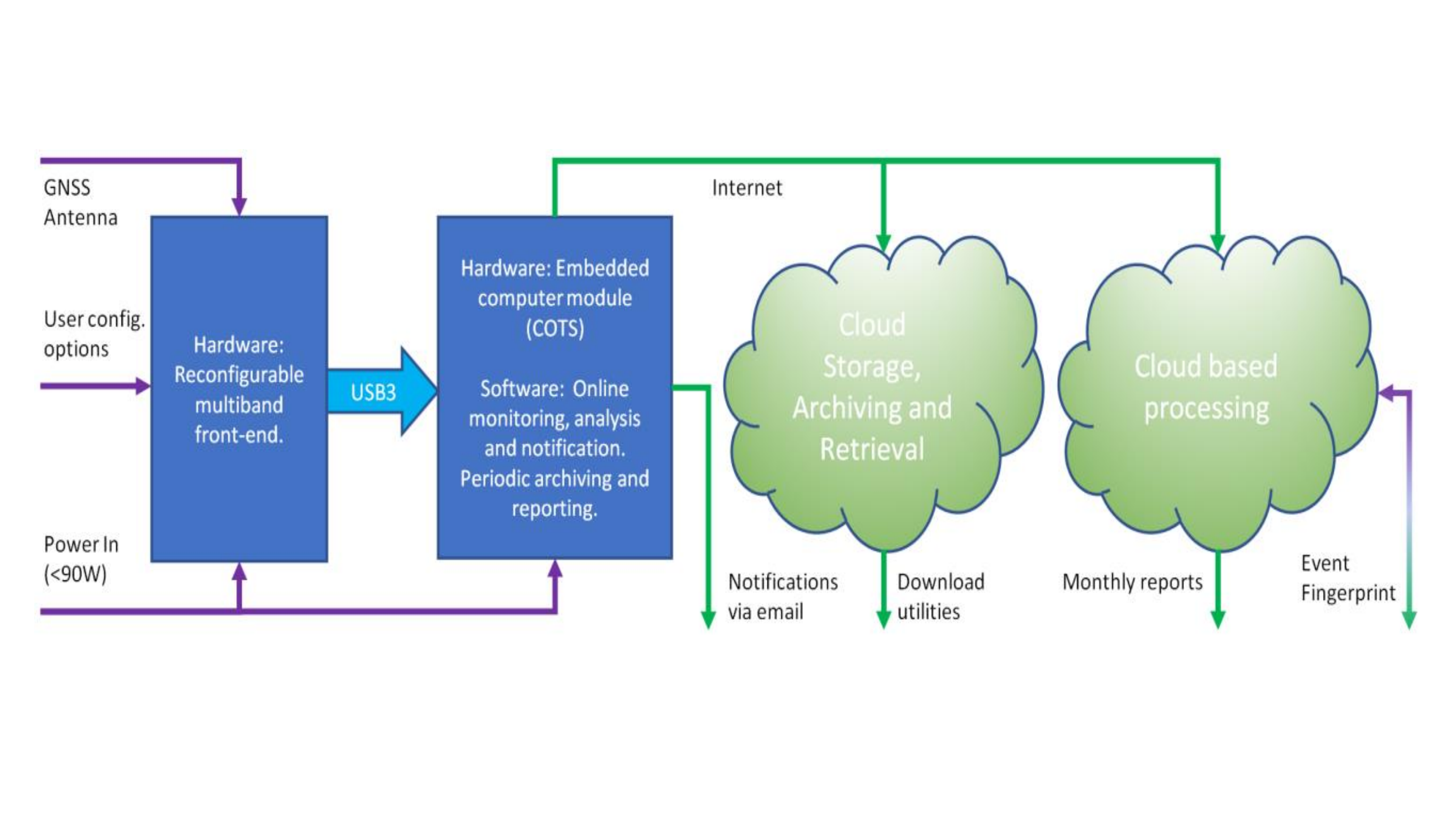}
	\caption{ARFIDAAS system design (\protect\citeSeminar{Sintef_Aiden}, presentation).}
	\label{fig:ARFIDAAS2}
\end{figure*}

\begin{figure}
\centering
\begin{subfigure}{.5\textwidth}
  \centering
  \includegraphics[clip, trim=4.5cm 0.0cm 0.0cm 0.0cm,width=1.2\linewidth]{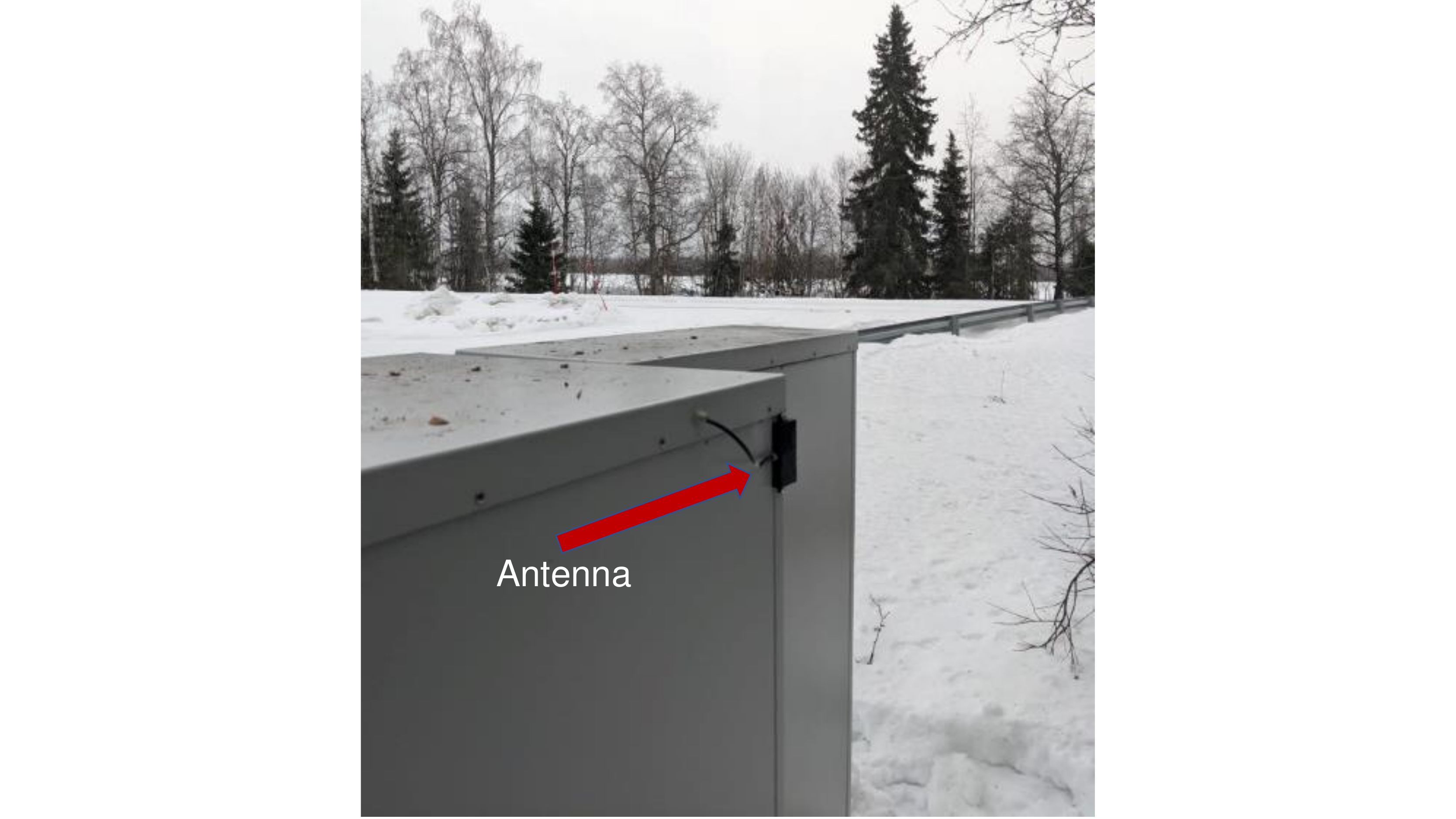}
  \caption{ARFIDAAS antenna.}
  \label{fig:sub1}
\end{subfigure}%
\begin{subfigure}{.5\textwidth}
  \centering
  \includegraphics[width=6.3cm,height=4.7cm ]{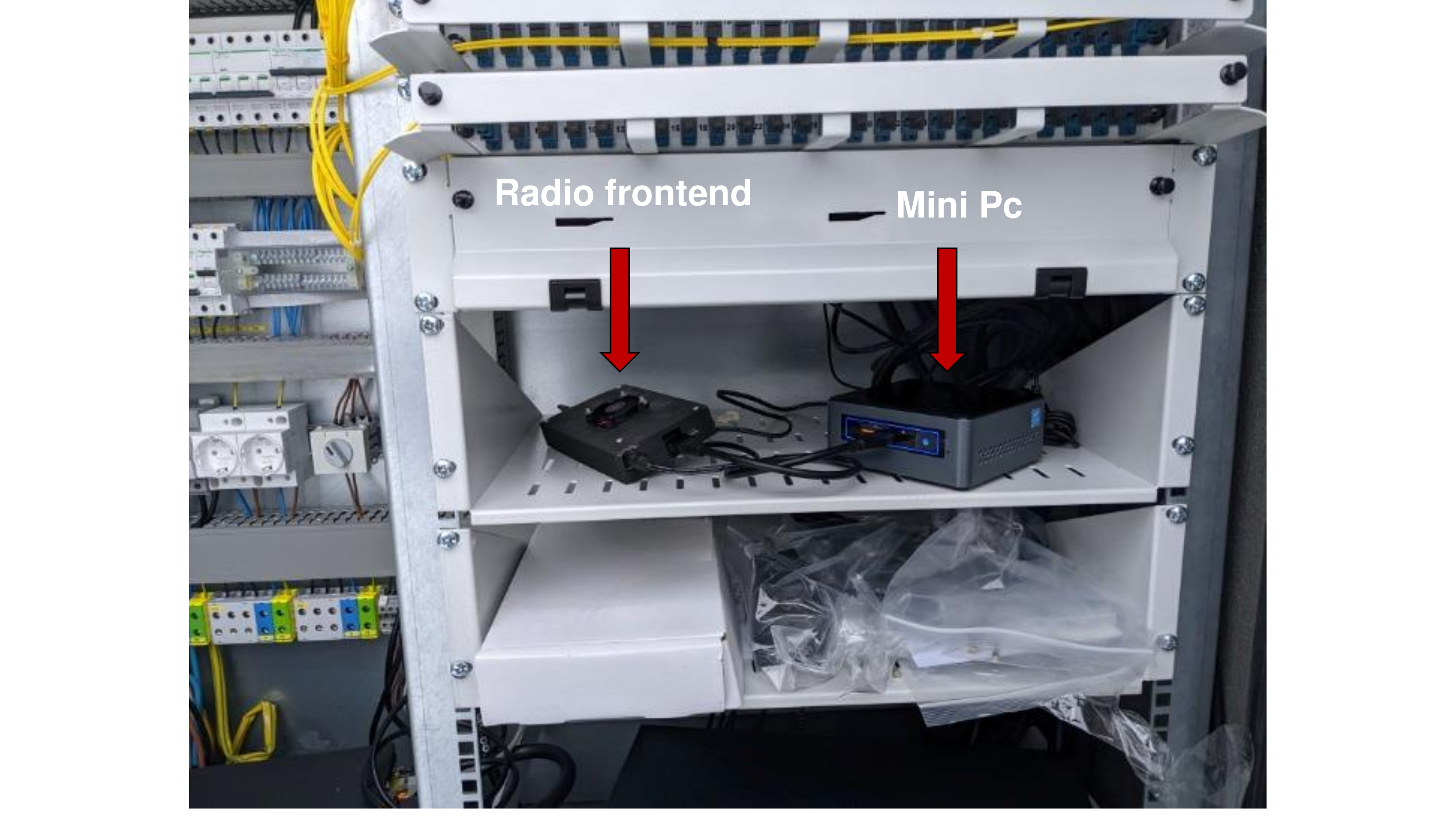}
  \caption{Inside an ARFIDAAS station.}
  \label{fig:sub2}
\end{subfigure}
\caption{ARFIDAAS monitoring station (\protect\citeSeminar{Sintef_Aiden}, presentation).}
\label{fig:ARFIDAASII}
\end{figure}

The Advanced RFI Detection, Analysis and Alerting System (ARFIDAAS II) project is a collaboration between SINTEF (Norway) and the University of Helsinki (Finland) with stakeholders NKOM (Norway) and Traficom (Finland). ARFIDAAS II is a European Space Agency NAVISP element 3 project running from 2021 through 2022 and continuation of SINTEF’s ARFIDAAS project operational from 2018 through 2019. The project is focused on the capture, collection and classification of RFI events impacting GNSS L-band signals and fingerprinting the jamming devices causing the RFI. One of the key features of the ARFIDAAS project is that it utilises custom monitoring hardware front-ends to simultaneously observe 240 MHz of aggregate spectrum divided into four tuneable sub-bands. The typical configuration is of one covering the L1 band including BeiDou B1 through GLONASS G1 signals, and the other three partially overlapping bands spread between the Galileo E5a+E5b, GPS L2 and Galileo E6 signals. The custom developed software for triggering, analysis and reporting runs on an off the shelf low cost computer. One of the main drivers in the front-end development was low deployment cost. At present, the cost of the whole system (Figure \ref{fig:ARFIDAAS2}); front-end, computer, cabling and an antenna (Figure \ref{fig:ARFIDAASII}), is below 2000 euros (\citeSeminar{Sintef_Aiden}, presentation). Low costs would enable increasing monitoring network density and thereby the detection of situations of potential failure in PVT computation.

 \begin{figure}[!t]
	\centering
	\includegraphics[clip, trim=0.0cm 0.0cm 0.0cm 0.0cm,width=1\columnwidth]{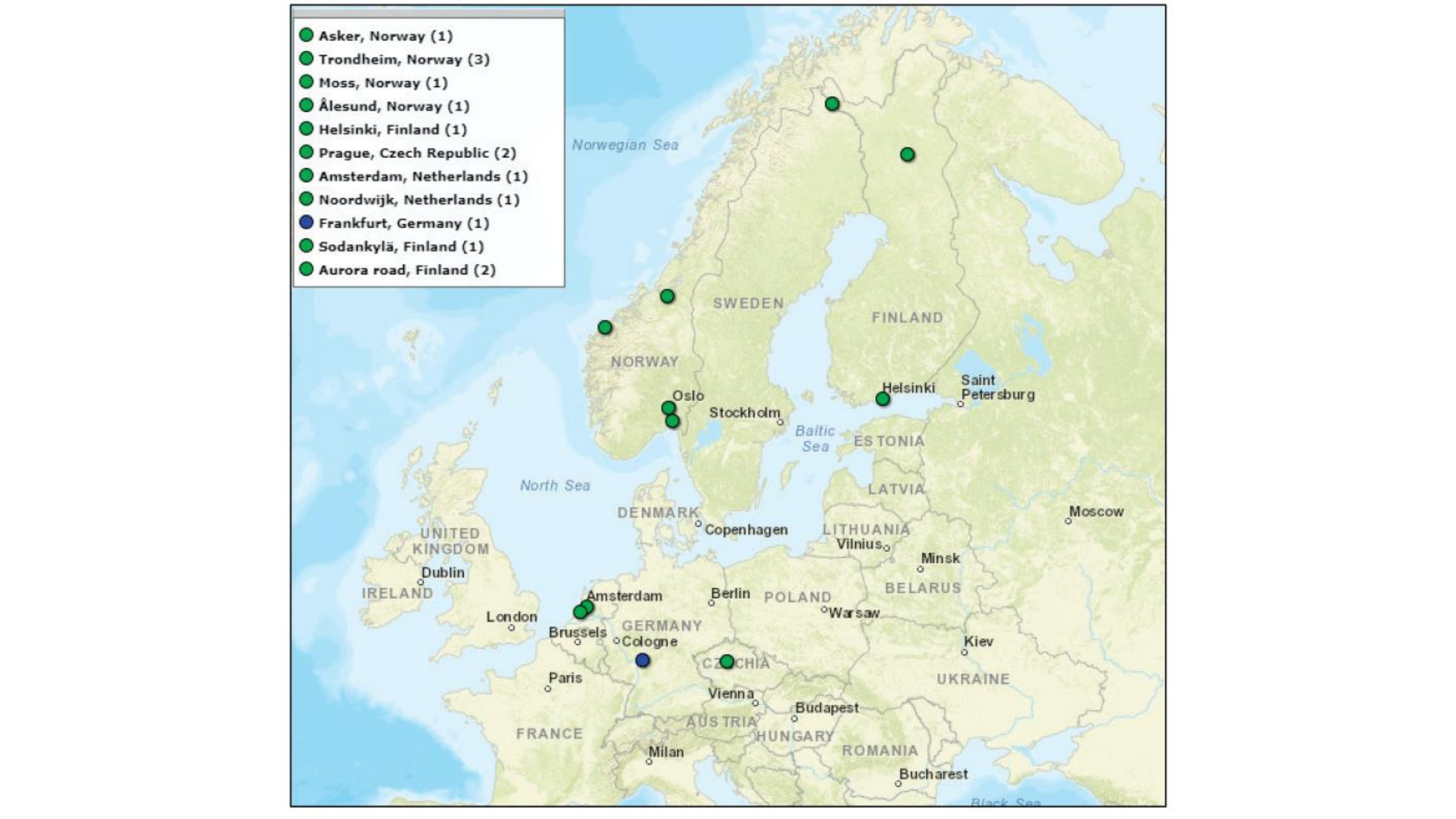}
	\caption{Locations of presently deployed ARFIDAAS systems in green, with expected 2022 
deployment sites in blue, and the number of systems indicated in brackets (\protect\citeSeminar{Sintef_Nadia}, presentation).}
	\label{fig:ARFIDAAS1}
\end{figure}

Between 2019 and 2021, eleven ARFIDAAS quad-band RFI monitoring stations had been deployed between Norway (6), Finland (1), the Czech Republic (2), and the Netherlands(2), and have together observed nearly five thousand RFI events impacting one or more GNSS signal bands (\citeSeminar{Sintef_Nadia}, presentation). At the beginning of 2022, three more stations were deployed to the Arctic environment in Northern Finland (\citeSeminar{UH_Laura}, presentation). During the five operational months, over ten RFI events have been observed. One more station is being deployed into Germany during 2022. The full geographical coverage of the stations is seen in Figure \ref{fig:ARFIDAAS1}, where the operating sites are shown with green and the planned 2022 deployment in blue, number of systems per site shown in brackets after each site name.






\section{Summary}

The disruptions to GNSS based navigation systems due to intentional and unintentional interference is a global phenomenon. Therefore, interference management techniques are crucial to mitigate interference effects. This white paper provides an overview of GNSS interference sources and methods to detect it. Specifically, traditional and machine learning methods are used to detect abnormal behaviour in the GNSS signal. Basic type of autoencoder is used for anomaly detection in the CWI interference type.
In addition, jammer fingerprinting methods are described. RF fingerprinting is based on detecting unique signal distortions caused by hardware irregularities, and this can be used to detect the signal transmitter. Moreover, methodologies to mitigate interference such as, Deeply coupling methods and Galileo Public Regulated Service (PRS) were discussed. Lastly, three GNSS signal monitoring infrastructures were briefly introduced.

\bibliographystyleSeminar{apalike}
\bibliographySeminar{new_bib}

\bibliographystyle{apalike}
\bibliography{bibliography}


\begin{table}[]
\caption{List of Abbreviations (alphabetical order)}
\label{tab:Abbrivations}
\begin{tabular}{ll}
\hline
Abbreviations & \\ \hline
AoA & Angle of Arrival \\
API & Application Program Interface\\
C/N0 & Carrier-to-Noise density ratio \\
CPA & Competent PRS Authority \\
CWI & Continuous Wave Interference \\
DGNSS & Differential-GNSS \\
DGPS & Differential GPS \\
ESA & European Space Agency \\
EU & European Union  \\
FDAF & Frequency Domain Adaptive Filter\\
FFT & Fast Fourier Transform \\
FGI &  Finnish Geospatial Research Institute \\
FinnRef & Finnish National Reference Network \\
GNSS  & Global Navigation Satellite Systems \\
HAS & High Accuracy Service \\
IGS & International GNSS Service \\
INS & Inertial Navigation System \\
LEO & Low Earth Orbit \\
LSTM & Long Short Term Memory \\
MEMS & Micro-Electro-Mechanical System \\
NEMA & National Marine Electronics Association \\
NLOS & Non Line Of Sight \\
NRTK & Network Real-Time Kinematic\\
OAPP & Orolia Academic Partnership Programs \\
OS & Open Service \\
PHM & Passive Hydrogen Maser \\
PRS & Public Regulated Service \\
PVT & Position, Velocity, and Time \\
RINEX & Receiver Independent Exchange \\
RHCP & Right-Hand Circular Polarization \\
RAFS & Rubidium Atomic Frequency Standard \\
RFI & Radio Frequency Interference \\
RIMS & Ranging and Integrity Monitoring Station \\
RTK & Real Time Kinematic \\ 
SAR & Search and Rescue Service \\
SDR & Software Defined Radio \\
SOI & Signal of Interest \\

\end{tabular}
\end{table}


\end{document}